\documentclass[aps,prl,twocolumn,showpacs,superscriptaddress]{revtex4}

\usepackage{graphicx}

\begin{document}

\preprint{12}

\title{Photo-Conductivity Response at Cyclotron-Resonance Harmonics in a Non-Degenerate 2D Electron Gas on Liquid Helium}
\author{R. Yamashiro}
\author{L.~V. Abdurakhimov}
\author{A.~O. Badrutdinov}
\affiliation{Okinawa Institute of Science and Technology, Tancha 1919-1, Okinawa 904-0495, Japan}
\author{Yu.~P. Monarkha}
\affiliation{Institute for Low Temperature Physics and Engineering, 47 Lenin Avenue,
61103, Kharkov, Ukraine}
\author{D. Konstantinov}
\email[E-mail: ]{denis@oist.jp}
\affiliation{Okinawa Institute of Science and Technology, Tancha 1919-1, Okinawa 904-0495, Japan}
\date{\today}

\begin{abstract}
We report the first observation of an oscillatory photo-conductivity response at the cyclotron-resonance harmonics in a non-degenerate 2D electron system formed on the free surface of liquid helium. The dc conductivity oscillations are detected for electrons occupying the ground surface subband. Their period is governed by the ratio of the microwave frequency to the cyclotron frequency. Theoretical analysis of the photo-response in a strongly interacting electron system indicates that the observation can be explained by an oscillatory correction to the electron distribution function that appears for a large inelastic relaxation time because of photon-assisted scattering.
\end{abstract}


\maketitle


\indent Two-dimensional (2D) electrons in a perpendicular magnetic field $B$
populate equidistant Landau levels, and, under certain conditions, can
resonantly interact with the microwave (MW) electric field of frequency $%
\omega $. Selection rules allow direct photon-induced transitions only
between adjacent states ($n^{\prime }-n=\pm 1$), as in the cyclotron
resonance (CR) at $\omega _{c}\rightarrow \omega $ ($\omega _{c}=eB/M_{e}c$
is the cyclotron frequency). However, photon-assisted scattering of
electrons from system disorder can lead to a remarkable photo-response in
electron transport at CR harmonics ($m\omega _{c}=\omega $), where $%
m=2,3,... $ is an integer. Magneto-oscillations in the dc resistivity and
conductivity of a 2D electron gas discovered in high-mobility GaAs/AlGaAs
heterostructures~\cite{ZudSim-2001,ManSme-2002,ZudDu-2003,YangZud-2003}
represent an outstanding example of such a photo-response. An important
feature of these oscillations is that they are governed by the ratio $\omega
/\omega _{c}$. In the vicinity of $\omega /\omega _{c}=m$ both the
resistivity and conductivity curves have asymmetrical shape with minima near
$\omega /\omega _{c}=m+1/4$. At low temperatures and strong enough power,
resistivity minima evolve into zero-resistance states (ZRS).

\indent Various theoretical mechanisms have been proposed to explain these
microwave-induced resistance oscillations (MIRO) and ZRS (for a review see~%
\cite{DmiMirPol-2012}), but the subject is still under debate. For two
most elaborated mechanisms of MIRO proposed for semiconductor systems
("displacement"~\cite{DurSacRea-2003,RyzSur-2003} and "inelastic"~\cite%
{DmiMirPol-2003,DmiVavAle-2005}), electron gas degeneracy is not a crucial
point. Therefore, the same kind of magneto-oscillations governed by $%
\omega/\omega_c$ potentially can appear in a nondegenerate 2D system of
surface electrons (SEs) on liquid helium. A different type of
microwave-induced magnetoconductivity oscillations is observed in this
system~\cite{KonKon-2009,KonKon-2010} when the inter-subband frequency $%
(\Delta_2-\Delta_1)/\hbar =\omega_{2,1}$ matches $\omega$ (here $\Delta_1$
and $\Delta_2$ are the energies of the ground and first excited surface
subbands, respectively). This phenomenon was explained~\cite%
{Mon-2011,Mon-2012,KonMonKon-2013} by nonequilibrium population of the
second surface subband which triggers quasi-elastic inter-subband scattering
against or along the driving force, depending on the relation between $%
\Delta_2-\Delta_1$ and the Landau excitation energy $m\hbar\omega_c$. It was
found that oscillations vanished if $\omega$ was substantially different
from $\omega_{2,1}$, which means that they are actually governed by the
ratio $\omega_{2,1}/\omega_c$.

The displacement mechanism of MIRO is based on the effect of the strong MW
field on in-plane electron scattering. A displacement of the electron orbit
center $X^{\prime}-X$ which follows from energy conservation for
photon-assisted scattering by impurities depends strongly on the relation
between $\hbar\omega$ and $\hbar\omega_c$. In the inelastic mechanism, MIRO
is caused by an oscillatory correction to the electron distribution function
which appears because of photon-assisted scattering to high Landau levels ($%
n^{\prime}-n=m$) and a large inelastic relaxation time. Theoretical analysis~%
\cite{Mon-2014} indicates that for electrons on liquid helium, the
photon-assisted scattering, which is important for both mechanisms, is
weaker than it is in GaAs/AlGaAs by the factor $M_e^*/M_e\sim 0.06$, where $%
M_e^*$ is the effective mass of electrons in GaAs/AlGaAs. Therefore,
observation of MIRO in a single subband of SEs on liquid helium requires
significantly higher MW power. This explains why MIRO, governed by the ratio
$\omega/\omega_c$, was not detected previously for electrons on liquid
helium.

\indent In this Letter, we report the first observation of $\omega/\omega_c$%
-periodic dc-magnetoconductivity oscillations induced by the MW in the
ground subband of a nondegenerate 2D electron system on the surface of
liquid $^4$He. In our experiment, the MW electric field polarized in the 2D
plane is significantly enhanced by employing a high Q-factor resonator. This
allows us to reach the necessary MW field sufficient to compensate for the reduction
in photon-assisted scattering of SEs caused by the free-electron mass $M_{e}$
and to produce a strong electron photo-response at CR harmonics. Results obtained are
compared with outcomes of the displacement and inelastic models of MIRO
applied to the nondegenerate strongly interacting 2D electron system on
liquid helium.

\begin{figure}[tbp]
\begin{center}
\includegraphics[width=8.5cm]{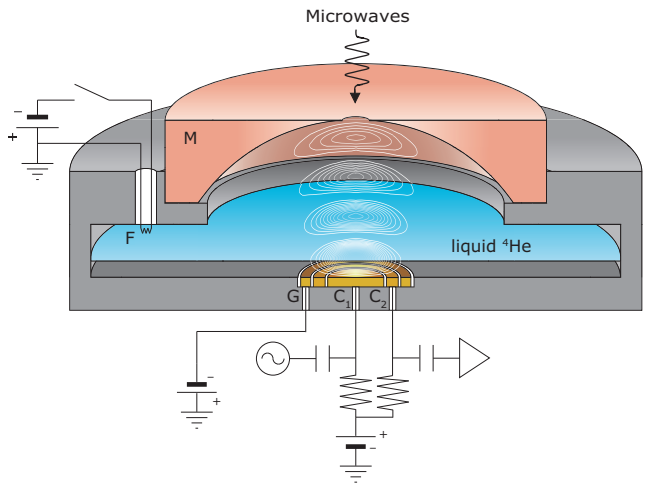}
\end{center}
\caption{(color online) Sketch of the experimental cell. White lines show
calculated electric field contours for $\text{TEM}_{003}$ mode. The pool of
2D electrons (not shown) is formed on the liquid surface above electrodes $%
\text{C}_1$ (central) and $\text{C}_2$ (middle). Further details are
provided in the text.}
\label{fig:1}
\end{figure}

\indent The sketch of our experimental cell is shown in Fig.~\ref{fig:1}.
The important feature of our experimental method is a semi-confocal
Fabry-Perot resonator operating in $\text{TEM}_{003}$ mode~\cite%
{PeltVas-2003}. The resonator is formed by two coaxial mirrors. The top
spherical (radius of curvature 9 mm) mirror M is made of high purity copper.
The bottom flat circular mirror (diameter 5.2~mm) is made of gold vaporized
onto a sapphire substrate and is divided into three concentric electrodes $%
\text{C}_1$, $\text{C}_2$ and $\text{G}$ by two gaps (width 0.01~mm) 1.5 and
2.13~mm from the center. Microwaves are produced by a broad-band
(75-110~GHz) source at room temperature, passed through an attenuator and a
low-pass filter (40~dB rejection above 139\,GHz), and guided into the
cryostat. There, microwaves are coupled from the waveguide to the resonator
through an aperture (diameter 1.3~mm) at the center of the mirror M. Below
1~K, $\text{TEM}_{003}$ mode has a frequency of 88.52~GHz and the quality
factor $Q\approx 900$. There are three nodes and four antinodes of the
electric field $E_{MW}$ between the mirrors. Calculated contours of $E_{MW}$
are plotted in Fig.~\ref{fig:1}. The fourth antinode, where $E_{MW}$ has the
global maximum value, is located about 1.0~mm from the flat mirror. The
waist of the Gaussian beam at this antinode is about 4.0~mm.

\indent The $^4$He gas is condensed into the cell and the liquid level is
set between the mirrors. During condensation, the resonant frequency of the
Fabry-Perot cavity, which is monitored by measuring the MW signal reflected
from the cavity, shifts due to a change in the dielectric constant of the
media filling the cavity. This allows us to set the liquid level at the
fourth antinode of $E_{MW}$. Electrons are produced by briefly heating a
tungsten filament $\text{F}$, while a positive bias is applied to electrodes
$\text{C}_1$ and $\text{C}_2$. The 2D electrons form a circular pool on the
liquid surface above the biased electrodes, and the areal density can be
approximately determined from the condition of complete screening of the
static electric field above the charged surface~\cite{KonKon-2010}. After
charging, a negative bias is applied to the guard electrode $\text{G}$ to
prevent charge leakage into and from the pool. The diagonal dc conductivity $%
\sigma_{xx}$ of the 2D electron liquid is measured by a standard capacitive
coupling method~\cite{SomTan-1971} using electrodes $\text{C}_1$ and $\text{C%
}_2$ as a Corbino disk. Conductivity values are extracted from experimental
data using a numerical procedure similar to that described previously~\cite%
{KonKon-2010}.

At $T<0.7\,\mathrm{K}$ SEs are predominantly scattered by surface excitations
of liquid helium (ripplons). Ripplons represent a sort of 2D phonons with
the capillary wave spectrum $\omega _{r,q}=\sqrt{\alpha /\rho }q^{3/2}$
(here $\alpha $ and $\rho $ are the surface tension and mass density of
liquid helium respectively). The electron-ripplon scattering is similar to
usual electron-phonon scattering in solids; it is described by the parameter
$U_{q}=V_{r,q}Q_{q}N_{q}^{1/2}$, where $V_{r,q}$ is the
electron-ripplon coupling~\cite{MonKon-book-2004}, $Q_{q}^{2}=\hbar q/2\rho
\omega _{r,q}$, and $N_{q}$ is the ripplon distribution function. The
momentum relaxation rate of SEs is determined by quasi-elastic, one-ripplon
scattering with long wavelength ripplons $q\sim 1/l_{B}$, which yields $%
N_{q}\simeq T/\hbar \omega _{r,q}$ and $\hbar \omega _{r,q}\ll \Gamma _{n}$,
where $l_{B}^{2}=\hbar c/eB$, and $\Gamma _{n}$ is the Landau level
broadening. Thus, under usual conditions $\hbar \omega _{c}\gg T$ and $%
\Gamma _{n}\ll T$, nearly all electrons occupy the lowest Landau level, 
which distinguishes SEs on liquid helium from electrons in GaAs/AlGaAs.

\begin{figure}[tbp]
\begin{center}
\includegraphics[width=8.5cm]{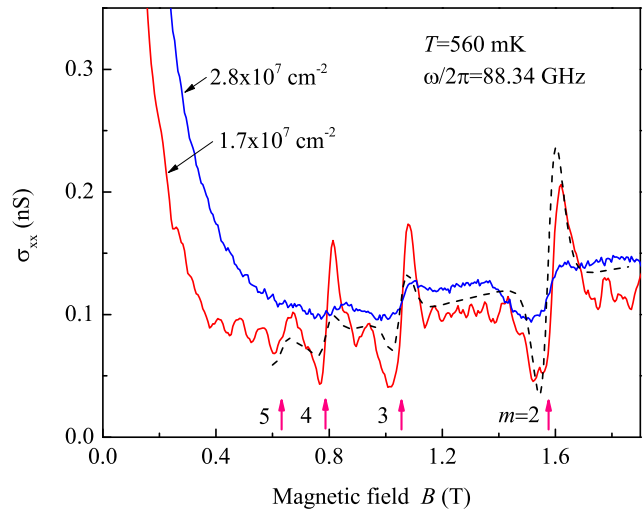}
\end{center}
\caption{(color online) $\protect\sigma_{xx}$ versus $B$ for surface
electrons in the Fabry-Perot resonator.
Data are taken at a fixed input MW
power,
and for two electron densities $%
n_s=2.8\times 10^7$ (blue) and $1.7\times 10^7$~cm$^{-2}$ (red). Arrows indicate
positions of $m=\mathrm{%
round}\left( \omega /\omega _{c}\right) $. Theoretical results obtained using the inelastic model
for the lowest $n_s$ are shown by the dashed line (black). }
\label{fig:2}
\end{figure}

\indent Figure~\ref{fig:2} shows $\sigma _{xx}$ versus $B$ for two values of
the electron density $n_{s}$. Data are taken at $T=0.56$~K and a fixed power
of $2.95$~mW measured at the output of the MW source. This corresponds to
the power incident on the resonator $P_{in}\approx 100$~$\mu$W. Higher source output power caused overheating of
the refrigerator; therefore it was avoided. Microwave frequency was adjusted
to excite resonant $\text{TEM}_{003}$ mode. According to our estimations,
almost all incident microwave power at resonance is dissipated inside the
cavity. For $P_{in}=100$~$\mu$W, we obtain the energy stored in the
resonator (approximately $10^{-13}$~J), which corresponds to the amplitude
of the MW electric field at the fourth antinode $E_{MW}\approx 10$~V/cm~\cite%
{KogLi-1966}. Oscillations of $\sigma _{xx}$ are clearly observed above 0.7~T for $%
n_{s}=2.8\times 10^{7}$~cm$^{-2}$. Oscillation amplitude strongly increases
with lowering electron density.

\indent It is important that we completely exclude the possibility that the
observed magnetooscillations originate from MW-induced electron transitions
to the first excited surface subband ~\cite{KonKon-2009,KonKon-2010} or from
transitions between higher excited subbands~\cite{KonKonJLTP-2010}. For the
first case, transitions appear only when the MW frequency matches the
intersubband transition frequency $\omega_{2,1}$, which can be Stark-tuned
by the electric (pressing) field applied normal to the surface. For
electrons on liquid $^4$He, this frequency is measured to be about 126~GHz
at zero pressing field~\cite{GrBr-1974}, and it increases with the pressing
field. Thus, in our experiment, the frequency of MW is at least 38~GHz lower
than the transition frequency $\omega_{2,1}$. This is overwhelmingly larger
than the corresponding transition linewidth which is about 10~MHz at $T=0.56$%
~K~\cite{ColLea-2002}. Moreover, the excitation of inter-subband resonance
requires the vertical component of $\mathbf{E}_{MW}$, which is significantly
reduced in the present setup due to employment of the $\text{TEM}$. Finally,
we find that the appearance of magnetoconductivity oscillations is
independent of the applied pressing field in the range at least 40~V/cm.
This corresponds to the upshift of inter-subband transition frequencies by
at least 30 GHz.

\begin{figure}[tbp]
\begin{center}
\includegraphics[width=8.5cm]{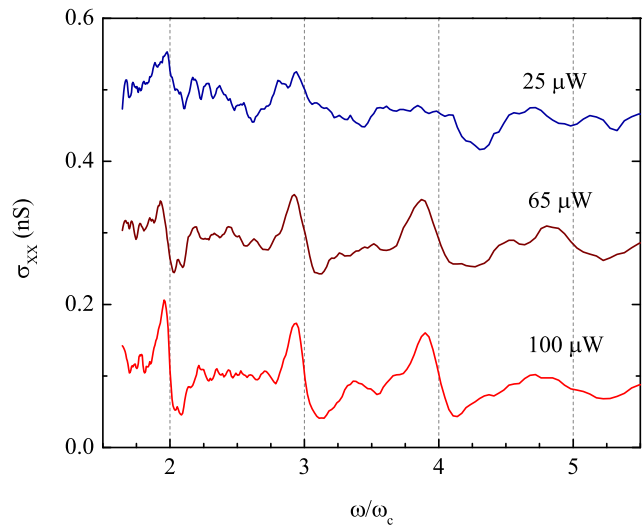}
\end{center}
\caption{(color online) $\protect\sigma _{xx}$ versus $\protect\omega /%
\protect\omega _{c}$ for $T=0.56$~K, $n_{s}=1.7\times 10^{7}$~cm$^{-2}$ and
different values of the incident MW power $P_{in}$. For clarity, curves for $P_{in}=25$
(dark blue) and 65~$\mu$W (brown) are upshifted by $0.4$ and $0.2\,\mathrm{nS}$
respectively.}
\label{fig:3}
\end{figure}

\indent Figure~\ref{fig:3} shows $\sigma _{xx}$ versus $\omega /\omega _{c}$
for $n_{s}=1.7\times 10^{7}$~cm$^{-2}$, $T=0.56$~K and several different
values of the incident power $P_{in}$. This graph proves that the observed oscillations are periodic
in $\omega /\omega _{c}$ and exhibit an asymmetric shape in the vicinity of
integer $\omega /\omega _{c}$, similar to MIRO in GaAs/AlGaAs
heterostructures (though data reported in Ref.~\onlinecite{ManSme-2002} differ from data of Refs.\onlinecite{ZudSim-2001,ZudDu-2003,YangZud-2003}). At $%
P_{in}=100$~$\mu$W, the amplitude of oscillations is about 50$\%$ of the "dark" value
of $\sigma _{xx}$.

The main oscillatory features shown in Figs.~\ref{fig:2} and \ref{fig:3} are in qualitative accordance
with the displacement model of MIRO applied to SEs on liquid helium~\cite%
{Mon-2014}, where the shape of oscillations is described by the derivative
of a Gaussian function $\partial G\left( \omega -m\omega _{c}\right)
/\partial \omega $ broadened due to interactions. When varying temperature
between $0.35\,\mathrm{K}$ and $0.56\,\mathrm{K}$, the amplitude of oscillations
approximately remains constant in spite of strong changes in
the electron-ripplon interaction ($U_{q}^{2}\propto T$), which also
agrees with the displacement model. Still, at maximum power used here,
the magnitude of oscillations given by
the theory can reach the observed values only if Coulomb interaction is
disregarded. In this case, the width of oscillatory features defined by the
averaged broadening $\Gamma _{0,m}=\sqrt{\left( \Gamma _{0}^{2}+\Gamma
_{m}^{2}\right) /2}$ is substantially smaller than in our data.
Coulomb forces induce additional broadening of the Gaussian $\Gamma
_{0,m}^{\ast }=\sqrt{\Gamma _{0,m}^{2}+x_{q}\Gamma _{C}^{2}}$, where $\Gamma
_{C}=\sqrt{2}E_{f}^{\left( 0\right) }l_{B}$,
$x_{q}=q^{2}l_{B}^{2}/2$ and $E_{f}^{\left( 0\right) }\simeq 3\sqrt{T}%
n_{s}^{3/4}$ is the typical fluctuational electric field acting on an
electron~\cite{DykKha-1979,FanDykLea-1997}. This makes the width of
oscillatory features comparable to the experimental data.
Naturally, the amplitude of oscillations diminishes and becomes an order of
magnitude smaller than that shown in Fig.~\ref{fig:2}.

Oscillatory features confined near $\omega /\omega _{c}=2$, $3$, and $4$ are well
separated; between them $\sigma _{xx}$ irregularly varies near "dark" values with an amplitude
practically independent of MW power.
This distinguishes oscillations observed here from MIRO in heterostructures,
and indicates the importance of Landau quantization for their origin.
At low $n_{s}$, the signal-to-noise ratio decreases due to a weak current, therefore measurements
at lower densities which might result in observation of ZRS become very challenging.
Hear we didn't study photo-conductivity response near
$\omega /\omega _{c}=1$ because under the CR condition already very low incident MW power
causes strong overheating of SEs which affects $\sigma_{xx}$.

Unfortunately, other theoretical mechanisms of MIRO discussed in the literature (reviewed in Ref.~\onlinecite{DmiMirPol-2012}) are developed for a degenerate
gas of noninteracting electrons and cannot be used for comparing with our
data. Extensions of these models applicable to a nondegenerate system of
strongly interacting electrons requires complicated studies which cannot be
presented in this work. In addition to the displacement model,
here we briefly discuss only the extension of the inelastic model, because
it is also associated with photon-assisted scattering.

The SCBA theory~\cite{AndUem-1974} applied to
electrons on liquid helium interacting with ripplons defines the effective
collision frequency~\cite{Mon-2014}

\begin{equation}
\text{\ \ }\nu =\frac{T}{\pi ^{2}n_{s}Ml_{B}^{4}}\sum_{n}r_{n,n}\int
d\varepsilon \left[ -\frac{\partial f\left( \varepsilon \right) }{\partial
\varepsilon }\right] g_{n}^{2}\left( \varepsilon \right) ,\text{ \ \ }
\label{e1}
\end{equation}%
where $f\left( \varepsilon \right) $ is the electron distribution function, $%
-g_{n}\left( \varepsilon \right) $ is the imaginary part of the
single-electron Green's function which represents the Landau level density
of states,
\begin{equation}
r_{n,n^{\prime }}=\frac{1}{\hbar T}\sum_{\mathbf{q}}U
_{q}^{2}x_{q}J_{n,n^{\prime }}^{2}\left( x_{q}\right) ,  \label{e2}
\end{equation}%
$J_{n,n^{\prime }}^{2}\left( x_{q}\right) $ is determined by $\left\vert
\left( e^{-i\mathbf{q}\cdot \mathbf{r}}\right) _{n,X,n^{\prime },X^{\prime
}}\right\vert ^{2}=\delta _{X,X^{\prime }+l_{B}^{2}q_{y}}J_{n,n^{\prime
}}^{2}\left( x_{q}\right) $. In Eq.(\ref{e1}) the overlapping of different
levels is neglected. For low levels, $g_{n}\left( \varepsilon \right) $ is
well described by a Gaussian function~\cite{Ger-1976}.

The MW field affects the electron
distribution function $f\left( \varepsilon \right) \simeq $ $f_{T}\left(
\varepsilon \right) +\tilde{f}\left( \varepsilon \right) $, where
$f_{T}\left( \varepsilon \right) \propto
e^{-\varepsilon /T}$, and $\tilde{f}\left( \varepsilon \right) $ is a
correction caused by photon-assisted scattering.
If the fraction of electrons excited to the level $%
n^{\prime }=m(B)\equiv \mathrm{%
round}\left( \omega /\omega _{c}\right) $ is small, a simple rate equation yields
\begin{equation}
\tilde{f}\left( \varepsilon \right) =\lambda \chi \frac{4f_{T}\left(
\varepsilon -\hbar \omega \right) g_{0}\left( \varepsilon -\hbar \omega
\right) r_{0,m}T}{\hbar \nu _{m}^{\left( 2r\right) }}.
\label{e3}
\end{equation}%
Here $\lambda =e^{2}E_{MW}^{2}/4M_{e}^{2}\omega ^{4}l_{B}^{2}$,
while $\chi $ depends only on $\omega /\omega _{c}$;
(it approaches $2$ if $\omega /\omega _{c}\rightarrow \infty $).
The inelastic relaxation rate $\nu
_{m}^{\left( 2r\right) }$ is caused by emission of pairs of short wavelength
ripplons~\cite{MonSokStu-2010} with $2\omega _{r,q}\simeq m\omega _{c}$.
The two-ripplon interaction is important only for iter-level relaxation, while
a quasi-uniform fluctuational field $E_f$ cannot lead to iter-level transitions
if $\Gamma _{C}\ll \hbar \omega _{c}$.

Eq.~(\ref{e3}) determines $\tilde{f}$ only for $\varepsilon $ near $\varepsilon _{m}$.
For noninteracting SEs, the shape of the "imprint" of the ground level density
$g_{0}\left( \varepsilon -\hbar \omega \right) $ is preserved because
transitions to higher levels occurs with $\hbar \omega _{r,q}\ll \Gamma_{n}$. The sharp maximum of
$g_{0}\left( \varepsilon -\hbar \omega \right) $ is the origin of
sign-changing terms in Eq.~(\ref{e1}). The new integrand factor $g_{0}\left(
\varepsilon -\hbar \omega \right) g_{n}^{2}\left( \varepsilon \right) $
selects levels nearest to the condition $\varepsilon
_{n}-\varepsilon _{0}=\hbar \omega $.
A fast drift velocity $\mathbf{u}_{f}$ of the electron orbit center in the
fluctuational field $\mathbf{E}_{f}$ leads to an inelastic correction in the
energy conservation $\hbar \mathbf{q}\cdot \mathbf{u}%
_{f}\simeq eE_{f}\left( X^{\prime }-X\right) $, and averaging over $\mathbf{E%
}_{f}$ results in additional broadening ($x_{q}^{1/2}\Gamma _{C}$) of the
maximum of $\tilde{f}\left( \varepsilon \right) $. Since backward
transitions caused by two-ripplon emission are strongly inelastic ($2\omega
_{r,q}\sim m\omega _{c}\gg \Gamma $), we assume that at the ground level $%
\tilde{f}\left( \varepsilon \right) $ is smoothed out and can be neglected
for low excitations.

In the inelastic model, the shape of $\sigma _{xx}$ variations is also
described by the derivative of Gaussian functions $\partial G\left( \omega
-m\omega _{c}\right) /\partial \omega $ broadened (slightly differently) due
to interactions. As compared to the displacement model, these variations has
an additional large factor $\sqrt{\pi }r_{m,m}T/\nu _{m}^{\left( 2r\right)
}\sqrt{\Gamma _{m}^{2}+x_{q}\Gamma_{C}^{2}}$.
The result of calculations performed for $n_{s}=1.7\times 10^{7}$~cm$^{-2}$,
$T=0.56$~K, and $E_{MW}= 2$~V/cm is shown in Fig.~\ref{fig:2} by
the dashed line. For high $B$, oscillatory
features are confined near $\omega /\omega _{c}=m$ which agrees with our
data. At low $B$, the broadening of terms with different $m$ increases and
they start overlapping. Then, the minima of $\sigma _{xx}$ approach the condition $%
\omega /\omega _{c}\left( B\right) \rightarrow m+1/4$. This
also agrees with positions of minima at large $m$ shown in Fig.~\ref{fig:3}.

It should be noted that the theoretical curve shown in Fig.~\ref{fig:2} is obtained for
a smaller amplitude of the MW field $E_{MW}= 2$~V/cm than that
estimated for our data.
If we include two-ripplon relaxation over intermediate levels ($n<m$), approximately
the same amplitude of oscillatory features is found at $E_{MW}= 3$~V/cm.
At $E_{MW}\sim 10$~V/cm, the assumption of small excitation used
above fails, and power saturation becomes important.
One can conclude that power saturation should restrict the amplitude of oscillatory features
especially those confined near small $m$. This explains rather weak power
dependence of the amplitude of oscillations which follows from Fig.~\ref%
{fig:3}. Also, one cannot exclude that small nonuniformity of
$\mathbf{E}_{f}$
can give an additional contribution to
the inelastic relaxation rate reducing $\tilde{f}$.
Still, experimental estimation of the inelastic relaxation rate
of SEs heated by the CR~\cite{Ede-1980} agrees with the result of the theory
of two-ripplon relaxation used in Eq.~(\ref{e3}).

In summary, for the first time in the nondegenerate 2D system of strongly interacting
electrons on liquid helium we observed MW-induced dc magnetoconductivity oscillations
governed by the ratio $\omega / \omega _{c}$ which are similar to those
discovered in heterostructures. This proves the universality of
the effect of MIRO, and potentially could help with identification of its
origin. A new feature observed for high magnetic fields is that oscillatory variations are strongly
confined near $\omega / \omega _{c}=2,3,$ and $4$; positions of minima are not fixed
to "magic" numbers $m+1/4$ which is in contrast with data obtained for GaAs/AlGaAs~\cite{ManSme-2002}.
We report also a new many-electron effect: the
amplitude of oscillations observed increases strongly with
lowering electron density, which is in agreement with the many-electron
treatment of photon-assisted scattering. Preliminary analysis shows that the
observation can be explained by an oscillatory correction to the electron
distribution function caused by photon-assisted scattering (inelastic model)
affected by strong internal forces.

\indent This work was supported by an internal grant from Okinawa Institute
of Science and Technology (OIST) Graduate University. We thank S.~A.
Vasiliev for fruitfull discussions and V.~P. Dvornichenko for technical
support.


\end{document}